# Authenticated Multiparty Quantum Key Agreement for Optical-Ring Quantum Communication Networks

Li-Zhen Gao[1], Xin Zhang[2]*, Song Lin[3]*, Ning Wang[2] and Gong-De Guo[2]*

[1]College of Computer Science and Information Engineering, Xiamen Institute of Technology, Xiamen, China, [2]College of Computer and Cyber Security, Fujian Normal University, Fuzhou, China, [3]Digital Fujian Internet-of-Things Laboratory of Environmental Monitoring, Fujian Normal University, Fuzhou, China

Quantum communication networks are connected by various devices to achieve communication or distributed computing for users in remote locations. In order to solve the problem of generating temporary session key for secure communication in optical-ring quantum networks, a quantum key agreement protocol is proposed. In the key agreement protocols, an attacker can impersonate a legal user to participate in the negotiation process and eavesdrop the agreement key easily. This is often overlooked in most quantum key agreement protocols, which makes them insecure in practical implementation. Considering this problem, the function of authenticating the user's identity is added in the proposed protocol. Combining classical hash function with identity information, we design the authentication operation conforming to the characteristics of quantum search algorithm. In the security analysis of the proposed protocol, quantum state discrimination is utilized to show that the protocol is secure against common attacks and impersonation attack. In addition, only single photons need to be prepared and measured, which makes our protocol feasible with existing technology.

## 1 INTRODUCTION

Communication is the exchange and transmission of information between people in a certain way. With the development of communication technology, people pay more attention to the privacy and security of data. In the present communication networks, RSA public key scheme is widely used for secure communication since it depends on the mathematical problem of large integer decomposition. However, the famous quantum factorization algorithm proposed by Shor [1] shows that this scheme is no longer safe. To ensure the security of communication, the research of quantum cryptography attracts people's attention. In contrast to the security of classical cryptography that are based on the assumption of computational complexity, the security of quantum cryptography relies on quantum-mechanics principles, which makes it unconditionally secure in theory. Since the first quantum key distribution protocol (BB84 protocol) was proposed [2], people try to solve some secure communication tasks with quantum cryptography, including quantum key distribution(QKD) [2–4], and quantum secure direct communication (QSDC) [5–7].

In addition to key distribution, key agreement (KA) is another major method of key establishment and plays a key role in the field of cryptography. In a key agreement protocol, two or more users in communication networks can agree on temporary session keys to achieve secure communication. As a significant cryptographic primitive, key agreement is flexibly used in multiparty secure computing, access control, electronic auctions, and so on. However, as the concept of quantum computer was put forward, classical key agreement was found to be as vulnerable to quantum computation as classical key distribution. Therefore, quantum key agreement (QKA) has been naturally proposed and has recently become a new research hotspot.

In 2004, [8] proposed the first two-party QKA protocol, which was designed based on the correlation of measurement results of EPR pairs. Unfortunately, this protocol is insecure, as shown in Ref. [9]. That same year, [10] proposed a fair and secure two-party QKA protocol based on BB84. Afterwards, researchers expands the number of negotiators from two to multiple parties to fit the actual scenarios. [11] proposed the first multi-party QKA (MQKA) protocol with Bell states and entanglement exchange in 2013. But in the same year, [12] pointed out that the protocol was unfair, then proposed a new MQKA protocol with single photons. Later [13] introduced two unitary operations and proposed the circle-type MQKA to improve the execution efficiency. Since then, many scholars have used various properties of quantum mechanics to design a few subtle MQKA protocols [14,15].

Actually, these protocols are only theoretically secure. Once they are used in practice, they will inevitably encounter the same problem as classical key agreement, namely, the impersonation attack. That is, an attacker may impersonate a legal user to participant in the protocol. Moreover, in classical key agreement protocols [16–19], the authentication of users is usually considered to protect against this particular attack. However, this is often overlooked in QKA. Although in some MQKA protocols, authentication of classical channels has been required to prevent classical messages from being tampered, message authentication is different from identity authentication. Therefore, in designing a secure QKA protocol, the authentication of users should be considered as in other authenticated quantum cryptographic protocols [20–24].

In this paper, an authenticated MQKA protocol for optical-ring quantum networks is proposed. The result shows that when all users perform the protocol honestly, they can get the correct negotiation key simultaneously. According to our analysis, the protocol in the network is secure against both common attacks and impersonation attacks.

## 2 PRELIMINARIES

### 2.1 Review of Communication Network

Let us start with a brief review of quantum communication networks. A communication network is a data link in which isolated individuals share resources and communicate through physical connections of various devices. The classical communication networks mainly consists of three parts: transmission, switching and terminal. According to the topological structure, it can be divided into bus, star, tree, ring and mesh types. Evidently, different types of networks are flexibly applied to different scenarios. This also provides the foundation for the study of quantum communication networks.

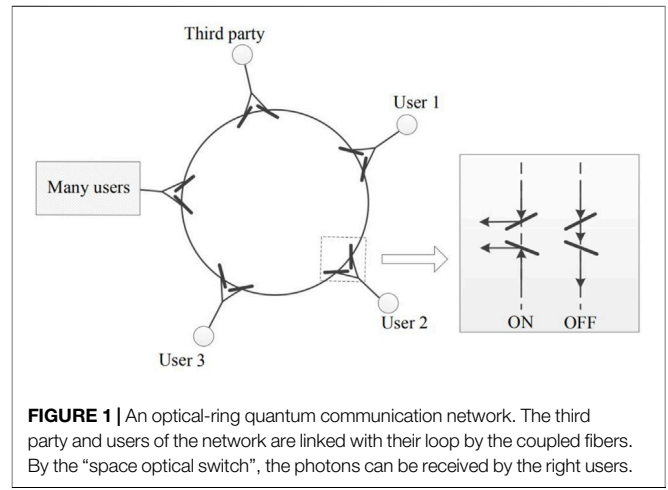

FIGURE 1 | An optical-ring quantum communication network. The third party and users of the network are linked with their loop by the coupled fibers. By the "space optical switch", the photons can be received by the right users.

Similar to classical communication networks, quantum communication networks can be classified into four types in terms of topology, which are: a passive-star network, an optical-ring network, a wavelength-routed network, and a wavelength-addressed bus network [25–27]. Among them, since the optical-ring topology is lower cost in the construction of the network, it is more conducive to promotion and studied by more people. In 2002, [28] have proposed an efficient multiuser quantum communication network, which can realize the QKD between arbitrary two users in the cascaded loop local networks. Inspired by them, we propose a multiparty quantum key agreement protocol in an optical-ring network.

Unlike the scheme of [28], the communication network in this paper only considers one loop, not a cascade. As shown in **Figure 1**, the network consist of three parts. 1) A third party. The third party is to facilitate communication between users on the network. 2) Users. Linked via coupled fibers and distributed in the communication network. 3) Switch. At each node, there is a "space optical switch", which is usually closed. Whenever a session key is required to be established, the photons are transmitted through the optical fiber among all users.

### 2.2 Review of Quantum Search Algorithm

Let us introduce Grover's search algorithm [29], which is used in the protocol. Suppose that we want to search a target $|\varphi_{mn}\rangle = |mn\rangle$, $m, n \in \{0, 1\}$, in the database of a set of two-qubit states, i.e.,

$$|\tilde{\varphi}_{xy}\rangle = \frac{1}{2}(|0\rangle + (-1)^y|1\rangle)(|0\rangle + (-1)^x|1\rangle), \quad (1)$$

where $x, y \in \{0, 1\}$. In order to search the target, two specific unitary operations need to be performed on $|\tilde{\varphi}_{xy}\rangle$. Namely, the phase reversal operation $U_{mn} = I - 2|\varphi_{mn}\rangle\langle\varphi_{mn}|$ and the amplitude amplification operation $V_{xy} = 2|\tilde{\varphi}_{xy}\rangle\langle\tilde{\varphi}_{xy}| - I$. After executing these two unitary operations, we get

$$V_{xy}U_{mn}|\tilde{\varphi}_{xy}\rangle = |\varphi_{mn}\rangle. \quad (2)$$

In this paper, since the global phase has no effect on results, it can be ignored.

In addition, the unitary operation $U_{mn}$ has two good properties, which have been used to design some quantum cryptographic protocols [30,31]. We suppose that a total of $r$ operations of $U_{mn}$ are performed on a two-qubit state. On the one hand, when the number of $r$ is odd, there is

$$U_{m_r n_r} \cdots U_{m_2 n_2} U_{m_1 n_1} = U_{mn}, \quad (3)$$

where $m = m_1 \oplus m_2 \oplus \cdots \oplus m_r$ and $n = n_1 \oplus n_2 \oplus \cdots \oplus n_r$, the symble $\oplus$ indicates bitwise Exclusive OR. In combination with **Eq. 2**, we know that the deterministic measurements can be obtained by single-particle measurement with basis $MB_Z = \{|0\rangle, |1\rangle\}$ at last. When the number of executions is even, there is

$$U_{m_r n_r} \cdots U_{m_2 n_2} U_{m_1 n_1} |\tilde{\varphi}_{xy}\rangle = |\tilde{\varphi}_{mn}\rangle, \quad (4)$$

where $m = x \oplus m_1 \oplus m_2 \oplus \cdots \oplus m_r$ and $n = y \oplus n_1 \oplus n_2 \oplus \cdots \oplus n_r$. Then, the measurements can be obtained by single-particle measurement with basis $MB_X = \{|+\rangle, |-\rangle\}$.

In our protocol, the user encodes his private input by unitary operation $U_{mn}$. In addition, to assure that the protocol satisfies the characteristics of Grover's algorithm after identity encoding, we design the identity encoding operations as $U_{00}$ and $U_{01}U_{10}$ respectively. In the protocol, the user always encodes his identity information after private input encoding, that is, the encoded quantum state is $U_{00}U_{mn}|\tilde{\varphi}_{xy}\rangle$ or $U_{01}U_{10}U_{mn}|\tilde{\varphi}_{xy}\rangle$. Furthermore, there exists $U_{m_1 n_1}U_{m_2 n_2} = X_{m_1 \oplus m_2, n_1 \oplus n_2}$. So when the identity encode is $U_{00}$, the private input stays the same. Otherwise, on the basis of **Eq. 3**, $U_{01}U_{10}U_{mn} = U_{\bar{m}\bar{n}}$, where $\bar{m} = m \oplus 1, \bar{n} = n \oplus 1$, which means the private input is flipped once.

## 3 QUANTUM KEY AGREEMENT PROTOCOL WITH IDENTITY AUTHENTICATION

Now, let us describe the proposed quantum key agreement protocol for optical-ring quantum communication networks, which can realize the key negotiation between arbitrary $N$ users in the networks. In this network, a third party $P_0$ is semi-trusted, who can perform the operation $U_{mn}$. Suppose there are $M$ users in the network and any $N$ of them perform the quantum key agreement. That is, the switches for $N$ users are turned on at the proper time, while the switches for the remaining $M - N$ users are constantly off.

Without loss of generality, assume that the first $N$ users participant in the negotiation, denoted as $P_1, P_2, \ldots, P_N$. They can not only perform the operations $U_{mn}$ and $V_{xy}$, but also have the ability to prepare and measure single particles. They want to negotiate a session key $K$ with the help of $P_0$, where $K = S_1 \oplus S_2 \oplus \cdots \oplus S_N$, and $S_i$ is $P_i$'s private input with length of $2n$. Furthermore, each user has an identity information $ID_i$ of length $l$. In order to ensure the legitimacy of these users' identity, it is necessary for $P_i(i = 1, 2, \ldots, N)$ to complete the identity authentication with $P_0$, who shares master key $\bar{k}_i$ with $P_i$.

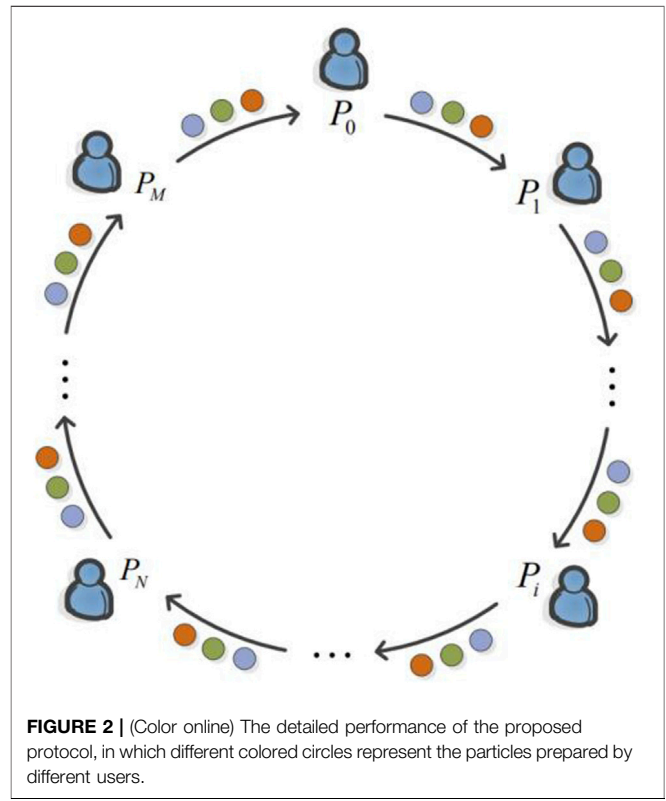

**FIGURE 2** | (Color online) The detailed performance of the proposed protocol, in which different colored circles represent the particles prepared by different users.

It should be noted that the switches for $P_N, \ldots, P_M$ are always closed, i.e., photons can be transmitted directly from $P_N$ to $P_0$. The general process of this protocol is shown in **Figure 2**.

In this quantum communication network, multi parties are connected by a quantum channel and a classical public channel. The quantum channel consists usually of an optical fiber. The classical channel, however, can be any communication link. Users and the third party can send classical messages via the classical channel, and these messages cannot be tampered with by anyone. That is, this transmitted classical message is required to be authenticated. Typically, the public classical channel can be achieved by broadcasting. However, it is worth noting that message authentication is different from identity authentication. So, we still need to verify the identity of each user. In the following, a description of the procedure for the protocol is given.

**Step 1**: $P_0$ and $P_i(i = 1, 2, \ldots, N)$ generate a random sequence $r_0$ and $r_i$ respectively and declare them through the classical channel. Then $P_0$ selects a hash function $f: 2^* \to 2^n$ and declares it. $P_0$ and each user calculate their authenticated message $h_i = f_{\bar{k}_i}(ID_i \parallel r_i \parallel r_0)$, where $\parallel$ denotes string concatenation.

**Step 2**: Each user $P_i(i = 1, 2, \ldots, N)$ generates a random bit sequence $L_i = (l_{i,1}, l_{i,2}, \ldots, l_{i,2n})$ and $B_i = (b_{i,1}, b_{i,2}, \ldots, b_{i,2n})$ with length of $2n$. In the process with $P_i$ as an initiator, an ordered sequence $T_i$ of $n$ two-qubit states is prepared by $P_i$ according to $L_i$:

$$T_i = \left(|\tilde{\varphi}_{l_{i,1}, l_{i,2}}\rangle, |\tilde{\varphi}_{l_{i,3}, l_{i,4}}\rangle, \ldots, |\tilde{\varphi}_{l_{i,2n-1}, l_{i,2n}}\rangle\right), \quad (5)$$

where, the $t$th quantum state is $|\tilde{\varphi}_{l_{i,2t-1}, l_{i,2t}}\rangle \in \{|\tilde{\varphi}_{00}\rangle, |\tilde{\varphi}_{01}\rangle, |\tilde{\varphi}_{10}\rangle, |\tilde{\varphi}_{11}\rangle\}$.

**Step 3**: $P_i$ performs the identity encoding operation on the $t$th quantum state of the sequence $T_i$ according to the values of $h_{i,t}$ and $b_{i,2t-1}$. Concretely, when $h_{i,t} \oplus b_{i,2t-1} = 0$, $P_i$ performs $U_{00}$; otherwise, he performs $U_{01}U_{10}$. Then the encoded sequence is denoted as $\tilde{T}_{i,i\boxplus 1}$, which is sent to the next user $P_{i\boxplus 1}$ over the quantum channel, where $\boxplus$ ($\boxminus$) represents addition (subtraction) module $N+1$.

**Step 4**: At this point, $P_{i\boxplus 1}$ opens his switch to receive the photons emitted by $P_i$. When $P_{i\boxplus 1}$ receives the sequence $\tilde{T}_{i,i\boxplus 1}$, he will perform corresponding operation to encode his own private input $s_{i\boxplus 1,2t-1}$, $s_{i\boxplus 1,2t}$. Namely, $P_{i\boxplus 1}$ performs the unitary operation $U_{s_{i\boxplus 1,2t-1},s_{i\boxplus 1,2t}}$ on the $t$th quantum state. After that, $P_{i\boxplus 1}$ performs his identity encoding similar to Step 3, and sends the encoded particle string $\tilde{T}_{i,i\boxplus 2}$ to the user $P_{i\boxplus 2}$.

**Step ($g+3$) ($g = 2, 3, \ldots, N$)**: Similar to Steps 4 and 3, $P_{i\boxplus g}$ encodes his private input and identity information. After that, he sends the encoded sequence $\tilde{T}_{i,i\boxplus(g+1)}$ to $P_{i\boxplus(g+1)}$. Notice that $P_0$ knows the hash values of all users. When the sequence of particles is transmitted to $P_0$, he calculate $h_{0,t} = \oplus_{i=1}^N h_{i,t}$. Based on the result, $P_0$ performs his identity encoding on the sequence. That is, if $h_{0,t}$ is 0, he performs $U_{00}$; Otherwise, he performs $U_{01}U_{10}$.

**Step ($N+4$)**: When all users $P_i (i = 1, 2, \ldots, N)$ receive the sequence $\tilde{T}_{i,i}$ from $P_{i\boxminus 1}$, they publish the random bit string $B_i$ in random order. Then, $P_i$ calculates $B^{2t-1} = \oplus_{i=1}^N b_{i,2t-1}$ ($t = 1, 2, \ldots, n$) and performs different operations as shown in follows.

1) When $B^{2t-1}$ is 0, $P_i$ measures the single photon with basis $MB_X$ directly to get the measurement result $|\tilde{\varphi}_{w_{i,2t-1},w_{i,2t}}\rangle$, then he can extract the session key:

$$k_{i,2t-1}k_{i,2t} = s_{i,2t-1}s_{i,2t} \oplus w_{i,2t-1}w_{i,2t} \oplus l_{i,2t-1}l_{i,2t}. \quad (6)$$

2) When $B^{2t-1}$ is 1, according to the classical bit sequence $L_i$, the unitary operation $V_{l_{i,2t-1},l_{i,2t}}$ is performed on the $t$th two-qubit state in the quantum sequence, then the particles are measured with basis $MB_Z$ to obtained the result $|\varphi_{w_{i,2t-1},w_{i,2t}}\rangle$. The agreement key is extracted as

$$k_{i,2t-1}k_{i,2t} = s_{i,2t-1}s_{i,2t} \oplus w_{i,2t-1}w_{i,2t} \oplus 11. \quad (7)$$

Obviously, each user $P_i$ can obtain the agreement keys $K_i = (k_{i,1}, k_{i,2}, k_{i,3}, k_{i,4}, \ldots, k_{i,2N-1}, k_{i,2N})$.

**Step ($N+5$)**: The eavesdropping detection process is executed. Namely, all users choose $\delta n$ samples to detect whether malicious or forged users exist. Specifically, each user $P_i$ randomly selects $\lfloor \frac{\delta n}{N} \rfloor$ samples from $K_i$, and declares these samples' positions. Then, he requires the other users $P_j (j \neq i)$ to announce the corresponding part of $K_j$. Since only legitimate users know the correct hash values and make the hash values satisfy $h_0 \oplus (\oplus_{i=1}^N h_{i,t}) = 0$, the users can get a consistent negotiation key by step ($N+4$). Afterwards, $P_i$ calculates the error rate according to his $k_{i,m}$ and the other users' $k_{j,m}$. That is, the number of inconsistencies in the sample as a proportion of the total sample size. If the error rate exceeds a certain threshold, the protocol is abandoned. Otherwise, the other users $P_j (j \neq i)$ perform similar actions. It should be noted that there are no common elements in the samples selected by all users. Finally, the remaining particles form their session key.

**TABLE 1** | The classical sequences of the example.

| | $P_1$ | $P_2$ | $P_3$ |
|---|---|---|---|
| $ID_i$ | 010,110 | 001,101 | 100,011 |
| $r_i$ | 1,100 | 1,111 | 0,010 |
| $h_i$ | 0,111 | 1,010 | 0,101 |
| $S_i$ | 01,101,011 | 01,000,100 | 10,110,001 |
| $L_i$ | 10,001,101 | 00,110,110 | 01,101,100 |
| $B_i$ | 00,100,100 | 10,110,010 | 11,011,110 |
| $B_{i,2t-1} \oplus h_{i,t}$ | 0,011 | 0,111 | 1,110 |

To illustrate the negotiation process more clearly, we give an example with ($N = 3$, $M = 5$). Similarly, we assume that $P_1$, $P_2$ and $P_3$ are involved in key negotiation and the switches for $P_4$ and $P_5$ are always off. In this case, $P_1$, $P_2$ and $P_3$ respectively hold secret inputs with length of 8 (i.e. $n = 4$), $S_1 = 01,101,011$, $S_2 = 01,000,100$, and $S_3 = 10,110,001$ and identity information with length of 6, $ID_1 = 010,110$, $ID_2 = 001,101$, $ID_3 = 100,011$. By the following steps, they can agree on a session key, $K = S_1 \oplus S_2 \oplus S_3$.

In Step 1, each user $P_i (i = 1, 2, 3)$ gets the random string $r_i$. In addition, $P_0$ generates $r_0$. Then, they can obtain the hash values $h_i$ according to the selected hash function. In the next step, $P_1 (P_2, P_3)$ generates two random 8-bit strings $L_1$ and $B_1$ ($L_2$ and $B_2$, $L_3$ and $B_3$). From $L_1(L_2, L_3)$, $P_1(P_2, P_3)$ prepares the single photons to obtain the two-particle sequence $T_1(T_2, T_3)$. The concrete values of these classical bit sequences are listed in **Table 1**.

After that, they proceed to the encoding phase of the protocol. The process with $P_1$ as the initiator is described in detail, where the sequence $T_1$ is back to $P_1$ after being encoded by all users. Concretely, in Step 3, $P_1$ performs the unitary operations $U_{00} \otimes U_{00} \otimes U_{01}U_{10} \otimes U_{01}U_{10}$ to encode his identity information. Afterwards, $P_1$ transmits the encoded sequence $\tilde{T}_{1,2}$ to $P_2$. When $P_2$ receives the sequence from $P_1$, he encodes his private input and identity information by performing unitary operations in Step 4, and sends to $P_3$. Similarly, $P_3$ also performs encoding operations. It is worth noting that the sequence $\tilde{T}_{1,0}$ sent from $P_3$ to $P_0$ passes through $P_4$ and $P_5$. When $P_0$ receives the sequence, he calculates $h_0 = 1,000$, which means his operations are $U_{01}U_{10} \otimes U_{00} \otimes U_{00} \otimes U_{00}$. After that, $P_0$ sends the encoded sequence to $P_1$. Obviously, the transmission process of particle sequences, which are prepared by $P_2$ and $P_3$, is similar to the above process. The variations of quantum states in three sequences are shown in **Table 2**. In Step 7, when all users receive the travelling particles $\tilde{T}_{1,1}, \tilde{T}_{2,2}, \tilde{T}_{3,3}$, they make the random strings $B_i$ public in random order. $P_1$ ($P_2$, $P_3$) calculates $B^{2t-1} = 0,010$. Therefore, $P_1$ ($P_2$, $P_3$) performs $I \otimes I \otimes V_{11} \otimes I$ ($I \otimes I \otimes V_{01} \otimes I$, $I \otimes I \otimes V_{11} \otimes I$). After that, they measure with appropriate measurement basis. By corresponding calculation, they get $K_1$, $K_2$, $K_3$ respectively. Apparently, if there is no eavesdropping, $K_1 = K_2 = K_3 = 10,011,110$.

## 4 ANALYSIS OF THE PROTOCOL

For a quantum key agreement protocol, it is generally required to satisfy correctness and security, regardless of the structure of the communication network. That is, all users can get the correct session key by executing the protocol. Security, on the other hand, implies that no attacker can obtain any information about the session key without being detected. Analysis shows that it can resist not only common external and internal attacks, but also impersonation attack.

**TABLE 2** | Change of the particle sequences during the encoding phase of the three-user protocol.

|   | $P_1$ | $P_2$ | $P_3$ |
|---|---|---|---|
| $T_i$ | $\|\tilde{\varphi}_{10}\rangle \otimes \|\tilde{\varphi}_{00}\rangle \otimes \|\tilde{\varphi}_{11}\rangle \otimes \|\tilde{\varphi}_{01}\rangle$ | $\|\tilde{\varphi}_{00}\rangle \otimes \|\tilde{\varphi}_{11}\rangle \otimes \|\tilde{\varphi}_{01}\rangle \otimes \|\tilde{\varphi}_{10}\rangle$ | $\|\tilde{\varphi}_{01}\rangle \otimes \|\tilde{\varphi}_{10}\rangle \otimes \|\tilde{\varphi}_{11}\rangle \otimes \|\tilde{\varphi}_{00}\rangle$ |
| $\bar{T}_{i,i\boxplus 1}$ | $U_{00}\|\tilde{\varphi}_{10}\rangle \otimes U_{00}\|\tilde{\varphi}_{00}\rangle \otimes \|\tilde{\varphi}_{00}\rangle \otimes \|\tilde{\varphi}_{10}\rangle$ | $U_{00}\|\tilde{\varphi}_{00}\rangle \otimes U_{00}\|\tilde{\varphi}_{00}\rangle \otimes \|\tilde{\varphi}_{10}\rangle \otimes \|\tilde{\varphi}_{01}\rangle$ | $\|\tilde{\varphi}_{10}\rangle \otimes \|\tilde{\varphi}_{01}\rangle \otimes \|\tilde{\varphi}_{00}\rangle \otimes U_{00}\|\tilde{\varphi}_{00}\rangle$ |
| $\bar{T}_{i,i\boxplus 2}$ | $U_{01}\|\tilde{\varphi}_{10}\rangle \otimes \|\tilde{\varphi}_{11}\rangle \otimes U_{10}\|\tilde{\varphi}_{00}\rangle \otimes U_{11}\|\tilde{\varphi}_{10}\rangle$ | $\|\tilde{\varphi}_{01}\rangle \otimes U_{00}\|\tilde{\varphi}_{00}\rangle \otimes U_{11}\|\tilde{\varphi}_{10}\rangle \otimes \|\tilde{\varphi}_{00}\rangle$ | $\|\tilde{\varphi}_{01}\rangle \otimes U_{00}\|\tilde{\varphi}_{01}\rangle \otimes U_{00}\|\tilde{\varphi}_{00}\rangle \otimes \|\tilde{\varphi}_{00}\rangle$ |
| $\bar{T}_{i,i\boxplus 3}$ | $\|\tilde{\varphi}_{10}\rangle \otimes U_{00}\|\tilde{\varphi}_{11}\rangle \otimes \|\tilde{\varphi}_{01}\rangle \otimes U_{10}\|\tilde{\varphi}_{10}\rangle$ | $\|\tilde{\varphi}_{00}\rangle \otimes \|\tilde{\varphi}_{00}\rangle \otimes \|\tilde{\varphi}_{01}\rangle \otimes U_{00}\|\tilde{\varphi}_{00}\rangle$ | $\|\tilde{\varphi}_{00}\rangle \otimes U_{10}\|\tilde{\varphi}_{01}\rangle \otimes \|\tilde{\varphi}_{01}\rangle \otimes U_{00}\|\tilde{\varphi}_{00}\rangle$ |
| $\bar{T}_{i,i}$ | $\|\tilde{\varphi}_{01}\rangle \otimes \|\tilde{\varphi}_{11}\rangle \otimes U_{00}\|\tilde{\varphi}_{01}\rangle \otimes \|\tilde{\varphi}_{00}\rangle$ | $\|\tilde{\varphi}_{11}\rangle \otimes \|\tilde{\varphi}_{10}\rangle \otimes U_{01}\|\tilde{\varphi}_{01}\rangle \otimes \|\tilde{\varphi}_{00}\rangle$ | $\|\tilde{\varphi}_{01}\rangle \otimes \|\tilde{\varphi}_{00}\rangle \otimes U_{10}\|\tilde{\varphi}_{01}\rangle \otimes \|\tilde{\varphi}_{11}\rangle$ |

## 4.1 Correctness

Obviously, with the example of three users in previous section, we can easily know that the session keys obtained by all users are equal. In this section, we will give a more rigorous proof to give a more convincing conclusion.

Without loss of generality, the session key derived from the $t$th quantum state is taken as an example. That is, we discuss whether or not the following equation holds:

$$k_{1,2t-1}k_{1,2t} = k_{2,2t-1}k_{2,2t} = \cdots = k_{N,2t-1}k_{N,2t}. \quad (8)$$

In the protocol, in order to obtain $k_{1,2t-1}k_{1,2t}$, $P_1$ prepares the initial quantum state $|\tilde{\varphi}_{l_{1,2t-1},l_{1,2t}}\rangle$ in Step 2. After that, $P_1$ and the other users perform their encoding operations on that quantum state in turn. For the sake of simplicity, let the identity of $P_i$ be $O_{i,t}$, where, $O_{i,t} = h_{i,t} \oplus b_{i,2t-1}$ when $i \neq 0$; $O_{0,t} = h_{0,t} = \oplus_{j=1}^{N} h_{j,t}$ when $i = 0$. Thus, in step $(N + 4)$, the quantum state received by $P_1$ is in:

$$U_{O_{0,t}}U_{O_{N,t}}U_{s_{N,2t-1},s_{N,2t}}\cdots U_{O_{3,t}}U_{s_{3,2t-1},s_{3,2t}}U_{O_{2,t}}U_{s_{2,2t-1},s_{2,2t}}U_{O_{1,t}}|\tilde{\varphi}_{l_{1,2t-1},l_{1,2t}}\rangle. \quad (9)$$

In the protocol, when $O_{i,t} = 0$, $U_{O_{i,t}} = U_{00}$; when $U_{O_{i,t}} = 1$, $U_{O_{i,t}} = U_{01}U_{10}$. Then, the parity times of the unitary operations $U_{xy}$ performed by the users are

$$C = O_{1,t} \oplus O_{2,t} \oplus \cdots \oplus O_{N,t} \oplus O_{0,t}. \quad (10)$$

By calculation, we get

$$C = \oplus_{i=1}^{N} b_{i,2t-1}. \quad (11)$$

Obviously, $C = B^{2t-1}$. So, in the protocol, the users can get the number of operations performed on the quantum state by calculating $B^{2t-1}$. Where, when $C = 0$, the unitary operation is executed an even number of times; otherwise, it is executed an odd number of times.

Due to the good reciprocity of the unitary operation $U_{xy}$, **Eq. 9** can be rewritten as:

$$U_{O_{0,t}}U_{O_{N,t}}\cdots U_{O_{3,t}}U_{O_{2,t}}U_{O_{1,t}}U_{s_{N,2t-1},s_{N,2t}}\cdots U_{s_{3,2t-1},s_{3,2t}}U_{s_{2,2t-1},s_{2,2t}}|\tilde{\varphi}_{l_{1,2t-1},l_{1,2t}}\rangle. \quad (12)$$

In addition, since $U_{xy}U_{xy} = I$, the identity encoding operation $U_{O_{0,t}}U_{O_{N,t}}\cdots U_{O_{3,t}}U_{O_{2,t}}U_{O_{1,t}}$ has the following conclusion. When $C = 0$, there are

$$U_{O_{0,t}}U_{O_{N,t}}\cdots U_{O_{3,t}}U_{O_{2,t}}U_{O_{1,t}} = I. \quad (13)$$

When $C = 1$, we get:

$$U_{O_{0,t}}U_{O_{N,t}}\cdots U_{O_{3,t}}U_{O_{2,t}}U_{O_{1,t}} = U_{11} \text{ or } U_{01}U_{10}. \quad (14)$$

So, **Eq. 12** is equivalent to

$$U_{s_{N,2t-1},s_{N,2t}}\cdots U_{s_{2,2t-1},s_{2,2t}}|\tilde{\varphi}_{l_{1,2t-1},l_{1,2t}}\rangle, \quad (15)$$

and

$$U_{11}U_{s_{N,2t-1},s_{N,2t}}\cdots U_{s_{2,2t-1},s_{2,2t}}|\tilde{\varphi}_{l_{1,2t-1},l_{1,2t}}\rangle, \quad (16)$$

or

$$U_{01}U_{10}U_{s_{N,2t-1},s_{N,2t}}\cdots U_{s_{2,2t-1},s_{2,2t}}|\tilde{\varphi}_{l_{1,2t-1},l_{1,2t}}\rangle. \quad (17)$$

Therefore, the user $P_1$ can perform different operations to extract the session key depending on the number of unitary operations.

As mentioned in Step $(N + 4)$ of the protocol, when the number of unitary operations is even, according to **Eq. 4**, $P_1$ directly measures the quantum state as in **Eq. 15** with $MB_X$ to obtain $|\tilde{\varphi}_{w_{1,2t-1},w_{1,2t}}\rangle$. From **Eq. 6**, we can obtain

$$\begin{aligned} k_{1,2t-1}k_{1,2t} &= s_{1,2t-1}s_{1,2t} \oplus w_{1,2t-1}w_{1,2t} \oplus l_{1,2t-1}l_{1,2t} \\ &= s_{1,2t-1}s_{1,2t} \oplus s_{2,2t-1}s_{2,2t} \oplus \cdots \oplus s_{N,2t-1}s_{N,2t}. \end{aligned} \quad (18)$$

When the number of unitary operations is odd, according to **Eqs. 2** and **3**, $P_1$ needs to perform the unitary operation $V_{l_{1,2t-1},l_{1,2t}}$ on the quantum state of **Eq. 16** or **Eq. 17**, and then use $MB_Z$ to obtain the result $|\varphi_{w_{1,2t-1},w_{1,2t}}\rangle$. In terms of **Eq. 7**, the agreement key is extracted as follows.

$$\begin{aligned} k_{1,2t-1}k_{1,2t} &= s_{1,2t-1}s_{1,2t} \oplus w_{1,2t-1}w_{1,2t} \oplus 11 \\ &= s_{1,2t-1}s_{1,2t} \oplus s_{2,2t-1}s_{2,2t} \oplus \cdots \oplus s_{N,2t-1}s_{N,2t}. \end{aligned} \quad (19)$$

Apparently, the agreement key is the sum of the private inputs of all users regardless of whether the number of operations is odd or even. Similarly, the quantum state $|\tilde{\varphi}_{l_{i,2t-1},l_{i,2t}}\rangle$ prepared by user $P_i$ is obtained after being encoded by other users as

$$U_{O_{i\boxminus 1,t}}U_{s_{i\boxminus 1,2t-1},s_{i\boxminus 1,2t}}U_{O_{0,t}}U_{O_{N,t}}U_{s_{N,2t-1},s_{N,2t}}\cdots U_{O_{i\boxplus 1,t}}U_{s_{i\boxplus 1,2t-1},s_{i\boxplus 1,2t}}U_{O_{i,t}}|\tilde{\varphi}_{l_{i,2t-1},l_{i,2t}}\rangle. \quad (20)$$

In the same way, the user $P_i$ can obtain

$$\begin{aligned} k_{i,2t-1}k_{i,2t} &= s_{i,2t-1}s_{i,2t} \oplus w_{i,2t-1}w_{i,2t} \oplus l_{i,2t-1}l_{i,2t} \\ &= s_{1,2t-1}s_{1,2t} \oplus s_{2,2t-1}s_{2,2t} \oplus \cdots \oplus s_{N,2t-1}s_{N,2t}. \end{aligned} \quad (21)$$

or

$$k_{i,2t-1}k_{i,2t} = s_{i,2t-1}s_{i,2t} \oplus w_{i,2t-1}w_{i,2t} \oplus 11 \\ = s_{1,2t-1}s_{1,2t} \oplus s_{2,2t-1}s_{2,2t} \oplus \cdots \oplus s_{N,2t-1}s_{N,2t}. \quad (22)$$

From **Eqs. 18** and **21** or **Eqs. 19** and **22**, it is shown that all users receive the same agreement key, i.e., **Eq. 8** holds. Therefore, the proposed protocol is correct.

## 4.2 Security

In this section, we analyze the security of the proposed protocol in the optical-ring quantum communication network. It not only proves that the protocol is secure against common external attacks and internal attacks, but also proves that impersonation attacks are also ineffective for this protocol.

### 4.2.1 External Attacks

Assuming Eve is an external attacker, who may try her best to eavesdrop on the private input $S_i$, the session key $K$ or the master key $\bar{k}_i$ without being detected. Next, we will discuss these three cases.

Case 1: Eavesdropping user's private input.

In the proposed protocol, each user has a private input. Since the private input constitutes the final session key, it is evident that it should be kept secret from others. Subsequently, we discuss that how Eve eavesdrops on the secret input of users.

During the process of the protocol, each user performs three operations: particle preparation, encoding private input and identity information, and single-particle measurement. Obviously, the disclosure of users' private inputs only occurs after the encoding operations. So, Eve's attacks mainly take place in the transmission of the particle sequence. In the following, we will consider two common attacks: intercept-resend attack and entangle-measure attack.

**Intercept-resend attack**. Eve firstly intercepts the particle sequence sent from $P_j$, and measures it. Based on the measurements, Eve re-prepares the sequence to send to $P_{j\boxplus 1}$. In this way, Eve hopes to obtain the private input without being detected. However, this is impossible. In the protocol, the carrier particles after different encoding operation numbers belong to two sets of non-orthogonal states, which are in

$$\{|++\rangle, |+-\rangle, |-+\rangle, |--\rangle\}, \quad (23)$$

or

$$\left\{\frac{1}{2}(|00\rangle - |01\rangle - |10\rangle - |11\rangle), \\ \frac{1}{2}(|00\rangle - |01\rangle + |10\rangle + |11\rangle), \\ \frac{1}{2}(|00\rangle + |01\rangle - |10\rangle + |11\rangle), \\ \frac{1}{2}(|00\rangle + |01\rangle + |10\rangle - |11\rangle)\right\}, \quad (24)$$

where, the second set can be converted into $\{|00\rangle, |01\rangle, |10\rangle, |11\rangle\}$ after the unitary operation $V_{xy}$. The identity encoding of user is determined by both hash values $h_j$ and random numbers $B_j$. Until step ($N + 4$) of the protocol, Eve does not know the value of $B_j$. Therefore, she can only perform random operations to obtain the information. That is, she randomly chooses the measurement base. Evidently, the probability that Eve selects a right measurement basis is approximately 50%. Then, a fake particle string is prepared and sent to $P_{j\boxplus 1}$ based on the measurement results. In this case, Eve introduces an error with a probability of $(\frac{1}{2} * \frac{3}{4}) = \frac{3}{8}$. Hence, this attack can be easily detected in step ($N + 5$). In a word, Eve cannot get user's private input without being detected in this way.

**Entangle-measure attack**. Assuming that Eve wants to perform the entangle-measure attack, she can intercept the travelling sequence prepared by $P_{j\boxplus 1}$, and apply entangling operation $U_E$ between her own ancillary particles and the intercepted particles. At last, she transmits the particles to $P_j$. $P_j$ just encodes his private input and identity information directly in the particles. Afterwards, the encoding sequence is transmitted to $P_{j\boxplus 1}$, at which point Eve intercepts again. Then, she measures the ancillary particles to infer the private input $S_j$.

Without loss of generality, the effect of Eve's unitary operation $U_E$ can be shown as

$$U_E|\alpha\rangle|E\rangle = |00\rangle|e_{00}\rangle + |01\rangle|e_{01}\rangle + |10\rangle|e_{10}\rangle + |11\rangle|e_{11}\rangle, \quad (25)$$

where, $|e_{00}\rangle, |e_{01}\rangle, |e_{10}\rangle, |e_{11}\rangle$ are pure states determined by $U_E$. The quantum state in the sequence intercepted by Eve again is shown in **Table 3**.

By simply calculating, we get the correlation between these eight states

$$|\alpha_7\rangle = |\alpha_0\rangle + |\alpha_4\rangle - |\alpha_3\rangle \\ = |\alpha_1\rangle + |\alpha_5\rangle - |\alpha_3\rangle \\ = |\alpha_2\rangle + |\alpha_6\rangle - |\alpha_3\rangle. \quad (26)$$

Obviously, there is a linear correlation between the quantum states after different coding operations. As Chefles and Barnett [32] said, the necessary and sufficient condition for distinguishing the quantum states is linear independence. Therefore, these linearly correlated quantum states cannot be unambiguous discriminated, which means Eve cannot obtain the private input $S_j$ through the entangle-measure attack.

Case 2: Eavesdropping the session key.

Here, we discuss whether Eve is able to eavesdrop on the session key $K$. Since $K = S_1 \oplus S_2 \oplus \cdots \oplus S_N$, Eve can generally use two methods to obtain the value of $K$. One is that Eve tries to eavesdrop each value of $S_i$ to infer the agreement key. However, from the analysis of Case 1, we know that Eve cannot succeed. The other method involves directly eavesdropping on the value of $K$. According to the analysis above, we know that Eve is unable to distinguish between two linearly correlated sets of quantum states. So, Eve will always be detected if the number of encoding operations is unknown. Then, what if she was directly involved in the protocol? Namely, she might execute the impersonation attack.

In this protocol, a semi-trusted third party, $P_0$, is introduced to help these parties accomplish this task. So, Eve may impersonates $P_0$ (called $\hat{P}_0$) to attack the protocol. In **Section 4.2.2**, we prove that a genuine $P_0$ cannot attain the session key. So, we could deduce directly that $\hat{P}_0$ is also unable to eavesdrop successfully. Hence, in this section, we focus on the second case. Namely, Eve wants to disguise herself as one user to execute the protocol with others. Suppose Eve impersonates $P_j$ (called $\hat{P}_j$). $\hat{P}_j$ prepares the quantum carriers, and hopes attain the $K_{-P_j}$, that is, the negotiation key of other users except $P_j$. In fact, this action can be detected by the authentication in step $(N+5)$. As $\bar{k}_j$ is only known to the valid user $P_j$ and $P_0$, $\hat{P}_j$ cannot calculate correct hash value $h_j$. Because of the special relationship between hash values, $h_0 = \oplus_{i=1}^N h_i$, as long as any one $h_j$ is error, this relationship is broken. Consequently, the quantum states will be changed, and the measurement using the measurement basis determined by $B^{2t-1}$ will result in random results. In other words, her impersonation was discovered. Therefore, the proposed protocol is secure against such attack.

**TABLE 3** | Quantum states after different encoding operations on the entangling state.

| | $xy \oplus mn$ | Encoded Quantum State |
|---|---|---|
| Odd times encoding $U_{xy}|\bar{\varphi}_{mn}\rangle$ | 00 | $|\alpha_0\rangle = \frac{1}{2}(|00\rangle|e_{00}\rangle - |01\rangle|e_{01}\rangle - |10\rangle|e_{10}\rangle - |11\rangle|e_{11}\rangle)$ |
| | 01 | $|\alpha_1\rangle = \frac{1}{2}(|00\rangle|e_{00}\rangle - |01\rangle|e_{01}\rangle + |10\rangle|e_{10}\rangle + |11\rangle|e_{11}\rangle)$ |
| | 10 | $|\alpha_2\rangle = \frac{1}{2}(|00\rangle|e_{00}\rangle + |01\rangle|e_{01}\rangle - |10\rangle|e_{10}\rangle + |11\rangle|e_{11}\rangle)$ |
| | 11 | $|\alpha_3\rangle = \frac{1}{2}(|00\rangle|e_{00}\rangle + |01\rangle|e_{01}\rangle + |10\rangle|e_{10}\rangle - |11\rangle|e_{11}\rangle)$ |
| Even times encoding $X_{xy}|\bar{\varphi}_{mn}\rangle$ | 00 | $|\alpha_4\rangle = \frac{1}{2}(|00\rangle|e_{00}\rangle + |01\rangle|e_{01}\rangle + |10\rangle|e_{10}\rangle + |11\rangle|e_{11}\rangle)$ |
| | 01 | $|\alpha_5\rangle = \frac{1}{2}(|00\rangle|e_{00}\rangle + |01\rangle|e_{01}\rangle - |10\rangle|e_{10}\rangle - |11\rangle|e_{11}\rangle)$ |
| | 10 | $|\alpha_6\rangle = \frac{1}{2}(|00\rangle|e_{00}\rangle - |01\rangle|e_{01}\rangle + |10\rangle|e_{10}\rangle - |11\rangle|e_{11}\rangle)$ |
| | 11 | $|\alpha_7\rangle = \frac{1}{2}(|00\rangle|e_{00}\rangle - |01\rangle|e_{01}\rangle - |10\rangle|e_{10}\rangle + |11\rangle|e_{11}\rangle)$ |

Case 3: Eavesdropping the master key.

We discuss whether it is possible for Eve to eavesdrop on the master key $\bar{k}_i$. $\bar{k}_i$ is shared only by users $P_i$ and $P_0$ and is related to the hash value $h_i$. In the proposed protocol, the public information is identity $ID_i$, random string $r_i$, and hash function $f$. Evidently, Eve cannot infer $\bar{k}_i$ from these public information. So, she wants to infer $\bar{k}_i$ from $h_i$. However, $P_i$ does not disclose $h_i$. In the protocol, users decide the identity operation with $h_i$ and $B_i$. Even if the user announces the random string in step $(N+4)$, Eve does not have access to any information about $\bar{k}_i$ for $B^{2t-1} = \oplus_{i=1}^{N-1} b_{i,2t-1}$, independent of $h_i$.

### 4.2.2 Internal Attacks

Compared with external attackers, internal parties have greater capacity since they are involved in the execution of the protocol. In the following, we will discuss some attacks of users.

Case 1: Dishonest users' collusion attacks.

In the case of a single dishonest user, even if he participates in the protocol, he cannot obtain the hash values of other parties from the information he knows. Therefore, he can be detected in step $(N+5)$ just like an external attacker. It is important to note that there is more than one dishonest user in a protocol, and the most serious case is only one honest user. Obviously, in this case, all dishonest users want to conspire to eavesdrop the private input of the only honest party and determine the final key $K$. If the protocol is secure in the extreme case, it is secure in others. Next, we will discuss this situation.

Since $P_0$ is semi-trusted in the protocol, he cannot conspire with others. When $P_j$ is the only honest one, other dishonest users will attack $P_0$ and $P_j$. For simplicity, let us take the example of three users ($N=3$, $M=3$). The users $P_1$ and $P_3$ in particular position are assumed to be dishonest, denoted as $P_1^*$ and $P_3^*$. The running process of the example is shown in **Figure 3**. The attack will be performed in the following ways.

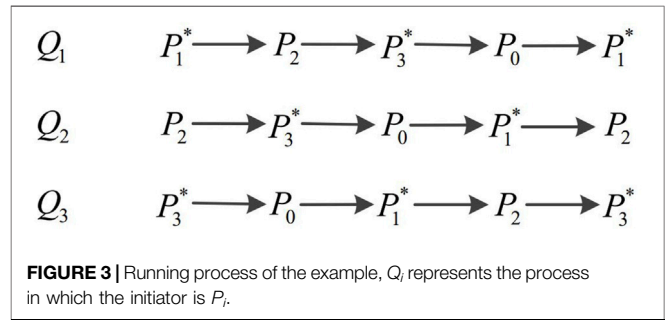

**FIGURE 3** | Running process of the example, $Q_i$ represents the process in which the initiator is $P_i$.

For one thing, we discuss the attack on $P_0$, where the dishonest users wish to obtain the hash value $h_0$. Through the above protocol, we know $h_0 = h_1 \oplus h_2 \oplus h_3$, which means it is determined by hash values of other users. In this case, since $P_1^*$ and $P_3^*$ conspired, they could know both $h_1$ and $h_3$. In spite of this, they could only know $h_1 \oplus h_3 = h_0 \oplus h_2$, but unable to determine the specific $h_0$ and $h_2$. Eventually, they have to obtain the information through the negotiation process. However, even $P_1^*$ and $P_3^*$ conspired, no matter what kind of attack they use, the deterministic information about $h_0$ could not be obtained. Because the encoded quantum states are nonorthogonal, they cannot be perfectly distinguished.

For another, we discuss the attack on $P_2$, in which the dishonest users wish to obtain the identity information $h_2$, the private input $S_2$, or determine the final key. Since $P_1^*$ and $P_3^*$ failed to attack $P_0$, it is impossible to determine whether the specific $h_2$ is 0 or 1. Next, we discuss attacks during the protocol process. Evidently, there is no information about $S_2$ is disclosed during $Q_2$ in **Figure 3**. In the encoding process $Q_3$, $P_2$ encodes his own information in the last stage of transmission. Since these three transmission processes are actually synchronized, it is obvious that $P_1^*$ and $P_3^*$ cannot encode the pre-negotiated message in $Q_3$ to determine the final key. So the most likely attack to obtain $S_2$ and determine the final key occurs in $Q_1$. In the transmission process of $Q_1$, $P_1^*$ prepares and encodes $n$ two-qubit particles, represented as $\tilde{T}_{1,2}$. Then, he sends them to $P_2$ and shares all his information with $P_3^*$. Since $P_3^*$ does not know whether $h_2 \oplus b_2$ is 0 or 1, he does not know whether the particles should be measured with basis $MB_X$ directly or basis $MB_Z$ after the operation $V$. Therefore, he can only get random results like an external attacker. To sum up, the proposed protocol is immune to this attack.

TABLE 4 | Comparison of several multi-party QKA protocols.

|  | Communication Network Type | Identity Authentication | Decoy Particles | Quantum Source | Particle Efficiency |
|---|---|---|---|---|---|
| [31] | Optical-ring | No | Yes | Single photon | $\eta = \frac{1}{(8N+1)N}$ |
| [15] | Optical-ring | No | No | Entangled particles | $\eta = \frac{1}{2N^2}$ |
| The proposed protocol | Optical-ring | Yes | No | Single photon | $\eta = \frac{2-\delta}{3N}$ |

Case 2: A semi-trusted third party's attack.

Here, $P_0$ is semi-trusted. That is, he cannot conspire with others, but misbehave on his own. For clarity, we represent the dishonest third party as $P_0^*$. $P_0^*$ wishes to obtain $P_j$'s private input $S_j$ or the session key $K$. Apparently, he has the advantage of knowing $h_j$. However, in this protocol, the user decides what kind of identity operation to perform through the value $h_j \oplus b_j$. Even if $P_0^*$ knows each person's hash value, he would not be able to get the correct operation information before step $(N+4)$. In addition, due to the non-cloning theorem, it is impossible for $P_0^*$ to preserve the quantum state without knowing about it. Therefore, this protocol is secure against the attack by a semi-trusted third party.

Case 3: A dishonest user's impersonation attack.

Users may also carry out impersonation attack in addition to the attacks described above. His purpose is to determine the agreement key by himself, and succeed in cheating others to accept this fake key. Even if $P_i$ is part of the protocol, he cannot perform correct identity encoding operation without knowing $P_j$'s hash value. Because the master key $\bar{k}_j$ is only shared by $P_j$ and $P_0$. Just like the impersonation attack by external attacker, the measurement result is random, which is difficult to pass the detection in step $(N+5)$. Therefore, a forged user cannot participate in the protocol and determine the session key without being detected.

Based on the above analysis, we prove that the protocol in the context of quantum networks is secure.

## 4.3 Efficiency

In this section, we will discuss the particle efficiency of the proposed protocol. According to [33], the particle efficiency is defined as

$$\eta = \frac{c}{q+b}, \quad (27)$$

where, $c$ is the length of the final shared key string, $q$ is the number of qubits transmitted in the quantum channel, and $b$ is the number of classical bits transmitted for decoding. In our scheme, the length of the final shared key is $(2-\delta)n$, the number of transmitted qubits is $n*N$, and the number of transmitted classical bits is $2n*N$. Therefore, the particle efficiency of the proposed protocol is

$$\eta = \frac{2-\delta}{3N}. \quad (28)$$

Without considering the detection of particles, the particle efficiency of the example presented in this paper is $\eta = \frac{2}{3*3} = 22.2\%$.

Compared with [31] and [15], although they both perform multi-user quantum key agreement in an optical-ring communication network, there are differences in the specific negotiation process. **Table 4** shows that our protocol is preferable as its readily accessible quantum resource, good security and high efficiency.

## 5 CONCLUSION

Before presenting our conclusion, we briefly discuss the hash function used in the protocol. In the protocol, we use the hash value to complete the authentication of the user. Only the legitimate user knows the correct hash value. In the absence of an impersonation attack, the hash values satisfy $h_0 = \oplus_{i=1}^N h_i$. So the selection criteria of the measurement basis is correct. Moreover, the hash values are not public. No one can obtain valid information from known information. Therefore, the introduction of classical hash function does not reduce the security of the protocol. Even if the classical hash function is corrupted by quantum computation, each user's master key is still secure. Since the master key is only shared by the third party and users through QKD, it can achieve absolute security. In addition, the security analysis shows that no matter what kind of attacks are used, the master key cannot be obtained by attackers. In this way, the shared master key can be reused and the user's identity can be authenticated, which greatly improves the practicability of the protocol.

In this paper, we study an authenticated quantum key agreement protocol, which is another main key establishment method in addition to quantum key distribution. This scheme enables key negotiation for any $N$ users in optical-ring quantum networks. Each user in the protocol has his own identity information and shares a master key with a semi-trusted third party. With the help of the third party, they can simultaneously obtain the negotiated key. Security analysis shows that the protocol is secure against common attacks and impersonation attack. Furthermore, the implementation of the protocol only requires preparing and measuring single particles, which can be easily implemented with current technology. And, our method can be easily applied to other MQKA protocols with authentication in quantum networks, so that they can resist impersonation attack in practical. Since the implementation of the protocol is inevitably affected by noise, the threshold value for the error rate should be provided before implementing it. As mentioned in [34], the exact value of the threshold is determined by a variety of practical elements, such as the desired level of security, the noise level of channels, etc. Therefore, choosing an appropriate threshold is complex, which is also the case for many multi-party quantum cryptographic protocols. Combined with quantum state discrimination, we will study this issue in the future.

## DATA AVAILABILITY STATEMENT

The original contributions presented in the study are included in the article/supplementary material, further inquiries can be directed to the corresponding authors.

## AUTHOR CONTRIBUTIONS

All authors listed have made a substantial, direct, and intellectual contribution to the work and approved it for publication.

## FUNDING

This work was supported by the National Natural Science Foundation of China (Grants Nos. 61,772,134, 61,976,053 and 62,171,131), Fujian Province Natural Science Foundation (Grant No. 2022J01186), the research innovation team of Embedded Artificial Intelligence Computing and Application at Xiamen Institute of Technology (KYTD202003), and the research innovation team of Intelligent Image Processing and Application at Xiamen Institute of Technology (KYTD202101).